\documentclass[10pt,conference]{IEEEtran}

\usepackage{subfigure,cite,graphicx,amsmath,amssymb,eufrak,mathrsfs,epsfig,dblfloatfix}
\usepackage{amsmath,amssymb,mathrsfs,graphicx}
\usepackage{psfrag}

\usepackage{cite}
\usepackage{graphicx,color,epsfig,rotating}
\usepackage{amsfonts,amsmath,amssymb,bbm}
\usepackage{algorithm,algorithmic}
\usepackage{subfigure}
\usepackage{amsmath}
\usepackage{cite}
\usepackage{mdwtab}
\usepackage{placeins}
\usepackage{psfrag, graphicx}
\usepackage[latin1]{inputenc}
\usepackage{amssymb}
\usepackage{multirow}
\usepackage{mathrsfs}

\long\def\comment#1{}

\newfont{\bbb}{msbm10 scaled 700}

\newfont{\bb}{msbm10 scaled 1100}

\newcommand{\PP}{\mbox{\bb P}}

\newcommand{\FF}{\mbox{\bb F}}

\newcommand{\EE}{\mbox{\bb E}}

\newcommand{\av}{{\bf a}}

\newcommand{\cv}{{\bf c}}

\newcommand{\ev}{{\bf e}}
\newcommand{\fv}{{\bf f}}
\newcommand{\gv}{{\bf g}}

\newcommand{\pv}{{\bf p}}
\newcommand{\qv}{{\bf q}}

\newcommand{\Am}{{\bf A}}

\newcommand{\Cm}{{\bf C}}

\newcommand{\Em}{{\bf E}}
\newcommand{\Fm}{{\bf F}}
\newcommand{\Gm}{{\bf G}}

\newcommand{\Wm}{{\bf W}}

\newcommand{\Dc}{{\cal D}}
\newcommand{\Ec}{{\cal E}}
\newcommand{\Fc}{{\cal F}}
\newcommand{\Gc}{{\cal G}}

\newcommand{\Uc}{{\cal U}}

\newcommand{\Vc}{{\cal V}}

\newcommand{\omegav}{\hbox{\boldmath$\omega$}}

\newcommand{\fsf}{{\sf f}}

\newcommand{\Fsf}{{\sf F}}

\renewcommand{\arg}{{\hbox{arg}}}

\newcommand{\eqdef}{\stackrel{\Delta}{=}}

\newcommand{\transp}{{\sf T}}

\newcommand{\be}{\begin{equation}}
\newcommand{\ee}{\end{equation}}
\newcommand{\bea}{\begin{eqnarray}}
\newcommand{\eea}{\end{eqnarray}}

\def\Fsf{ {\sf F}}

\def\fsf{ {\sf f}}

\newtheorem{defn}{Definition}
\newtheorem{theorem}{Theorem}

\newtheorem{lemma}{Lemma}

\newtheorem{corollary}{Corollary}

\begin{document}
\title{Caching and Coded Multicasting with Multiple Random Requests}
\title{Caching-Aided Coded Multicasting with Multiple Random Requests}



\author{\authorblockN{Mingyue Ji\authorrefmark{1},
Antonia Tulino\authorrefmark{2}\authorrefmark{3}, Jaime Llorca\authorrefmark{2}, and Giuseppe Caire\authorrefmark{1}}
\authorblockA{\authorrefmark{1} University of Southern California, CA, {\it email: \{mingyuej, caire\}@usc.edu}}
\authorblockA{\authorrefmark{2} Bell Labs, Alcatel-Lucent, NJ, {\it email: \{jaime.llorca, a.tulino\}@alcatel-lucent.com}}
 \authorblockA{\authorrefmark{3}EE Department, Universit\'a di Napoli Federico II, Italy.  {\it email: antoniamaria.tulino@unina.it}}
\vspace{-10mm}
}
\maketitle

\IEEEpeerreviewmaketitle

\begin{abstract}
The capacity of caching networks has received considerable attention in the past few years. 
A particularly studied setting is the 
\emph{shared link caching network}, in which a single source with access to a file library communicates with
multiple users, each having the capability to store segments (packets) of the library files, over a shared multicast link. 
Each user requests one file from the library according to a common demand distribution and the server sends a coded multicast message to satisfy all users at once. The problem consists of finding the smallest possible average codeword length to satisfy such requests. 
In this paper, we consider the generalization to the case where each user places $L\geq 1$ independent requests according to the same common demand distribution. 
We propose an achievable scheme based on random vector (packetized) caching placement and multiple groupcast index coding, shown to be order-optimal in the asymptotic regime in which the number of packets per file $B$ goes to infinity. 
We then show that the scalar  ($B=1$) version of the proposed scheme can still preserve order-optimality when the number of per-user requests $L$ is large enough. 
Our results provide the first order-optimal characterization of the shared link caching network with multiple random requests, revealing the key effects of $L$ on the performance of caching-aided coded multicast schemes.
\end{abstract}



\section{Introduction}  \label{intro}

Wireless data traffic has grown dramatically in the past few years and is expected to increase at an even faster pace in the near future, pushed by increasingly popular on-demand video streaming services~\cite{cisco13}. Such type of traffic is characterized by {\em asynchronous content reuse} \cite{6495773}, i.e., the fact that user demands concentrate on a relatively small set of files (e.g., 1000 titles of TV shows and movies), but the streaming sessions happen at arbitrary times such that {\em naive multicasting} of the same file (e.g., exploiting the inherent broadcast property of the wireless channel) yields no significant gain.   
An effective approach to leverage such asynchronous content reuse is {\em caching} content files at or close to the end users.  A significant amount of recent work has shown that caching at the wireless edge  yields very effective ways to trade-off expensive wireless bandwidth for relatively cheap memory at the wireless edge devices (e.g., access points or end user devices) 
\cite{llorca2013network, ji2013wireless, maddah2012fundamental,maddah2013decentralized,niesen2013coded,ji2015random,ji2014order2,ji2014average,shanmugam2014, ji2015Preserving, ji2015Efficient}. 


A particularly studied setting is the \emph{shared link caching network}, in which a single source with access to a library of $m$ files communicates with $n$  users, each with cache capacity $M$ files, via a shared multicast link. 
In \cite{maddah2012fundamental, maddah2013decentralized}, Maddah-Ali and Niesen studied the min-max rate for this network, i.e. the minimum codeword length needed to satisfy the worst-case users' demand. 
An achievable scheme based on a deterministic combinatorial cache construction for the placement phase and a linear coded multicast scheme for the delivery phase is proposed in \cite{maddah2012fundamental} and, through a cut-set information theoretic bound, it is shown to be order-optimal in the 
min-max sense. 
In \cite{maddah2013decentralized}, a simpler random uniform caching phase is shown to be sufficient for min-max rate order-optimality. 

In \cite{niesen2013coded,ji2015random}, the authors extended the results of \cite{maddah2012fundamental, maddah2013decentralized} to the case in which user demands follow a probability distribution. In \cite{niesen2013coded}, Niesen and Maddah-Ali presented a \emph{grouping} scheme based on applying the order-optimal min-max rate scheme separately within groups of files with similar popularity. However, 
missing coding opportunities between files in different groups and the fact that the file grouping did not take into account the joint effects of all the system parameters, 
yielded no order-optimal guarantees. 
In \cite{ji2015random}, Ji {\it et al.} presented a general random caching scheme based on a caching distribution designed according to the joint effects of all the system parameters and a chromatic-number based index coding delivery scheme that allows coding between any requested packet, shown to 
achieve average rate order-optimality under Zipf demand distributions. 

In this work, we consider the generalization of the setting in \cite{niesen2013coded, ji2015random} (one source, $m$ files, $n$ user caches, cache capacity $M$, one-hop multicast transmission, and random demands) when users make multiple simultaneous requests. 
This scenario may be motivated by a {\em FemtoCaching} network \cite{6495773} 
formed by $n$ small-cell base stations receiving data from a controlling ``macro'' base station via the cellular downlink. 
Each small-cell base station has a local cache of size $M$ file units and serves $L$ users through its local high-rate downlink. 
Hence, each small-cell base station can be thought of a user in our network model that makes $L$ file requests at once.
A similar network setting, but under worst-case demands, is addressed in \cite{maddah2013decentralized, ji2014order2}. 
The analysis in \cite{maddah2013decentralized} is constrained to the case in which the cache capacity $M$ scales linearly with the number of per-user requests $L$, and 
$m > nL$, 
which restricts the full characterization of the scaling effect of $n$ and $L$ in the system.
The work of \cite{ji2014order2} extends the result in \cite{maddah2013decentralized} providing the complete order-optimal characterization of the min-max rate for the shared link caching network with multiple per-user requests. 
In this paper, we address the more relevant demand setting in which users request files according to a popularity distribution with the goal of characterizing the order-optimal \emph{average} rate for the shared link caching network with multiple per-user random requests. 

Our contribution is two-fold. First, we generalize the random vector (packet based) caching placement and coded multicast scheme proposed in \cite{ji2014order2} to the case of multiple requests according to a demand distribution, where {\em multiple} means that each user makes $L \geq 1$ requests. The performance metric is the average number of equivalent file transmissions. We show that the proposed scheme is order-optimal under a Zipf demand distribution with parameter $\alpha$ in $[0,1)$. Second, by recognizing the effect of $L$ in the system, 
we introduce a random scalar caching placement scheme, i.e., caching entire files according to a probability distribution, and show that when $M$ and $L$ is sufficiently large, the order optimality of the shared link caching network can also be guaranteed. 

\section{Network model}

We consider a network with a single source node (server) connected to $n$ user nodes $\mathcal U = \{1, \cdots, n\}$ (caches) through a shared multicast link. 
The source has access to the whole content library $\mathcal{F} = \{1, \cdots, m\}$ containing $m$ files of equal size $F$ bits.
Each user node has a cache of size $M$ files (i.e., $MF$ bits).  
The shared link is a deterministic channel that transmits one file per unit time, such that all the users can decode the same multicast codeword. 
At each time unit (slot), each user requests a set of $L$ files in $\mathcal F$. Each request is placed independently according to a probability distribution $\qv = (q_1, \ldots, q_m)$, referred to as the
{\em demand distribution}. This is known a priori and, without loss of generality up to index reordering, 
has non-increasing components $q_1 \geq \cdots \geq q_m$. 
Such requests form a random matrix $\Fsf$ of size $L \times n$
with columns  $\fsf_u = [\fsf_{u,1}, \fsf_{u,2}, \cdots, \fsf_{u,L}]$ corresponding to the requests of each user $u \in \Uc$. The realization of $\Fsf$ is denoted as $\Fm = [\fv_1, \fv_2, \cdots, \fv_n]$, where $\fv_u = (f_{u, 1}, f_{u, 2} \ldots, f_{u, L})^{\rm T}$. 
The caching problem includes two distinct operations: the caching phase and the delivery phase. The caching phase (cache configuration) is done a priori, 
as a function of the files in the library, but does not depend on the request matrix realization $\Fm$. Then, during the delivery phase, at each time slot, given the current request matrix realization $\Fm$,  the source forms a multicast codeword and transmits it over the shared link such that all users can decode their requested files. 
Formally, we have: 

\begin{defn}
{\bf (Caching Phase)} The caching phase is a mappin of the file library $\Fc$ onto the user caches. Without loss of generality, 
we represent files as vectors over the binary field $\FF_2$. For each $u \in \Uc$, let $\phi_u: \FF_2^{mF} \rightarrow \FF_2^{MF}$ denote the caching function of
user $u$. Then, the cache content of user $u$ is given by $Z_u \triangleq \phi_u(W_f : f = 1, \cdots, m)$, where $W_f \in \FF_2^{F}$ denotes the $f$-th file in the library. 
\hfill $\lozenge$
\end{defn}

\begin{defn}
\label{def: Delivery Phase}
{\bf (Delivery Phase)} 
At each use of the network, 
a realization of the random request matrix $\Fm \in \Fc^{L \times n}$  is generated. 
The multicast encoder is defined by a fixed-to-variable encoding function 
$X : \FF_2^{mF} \times \Fc^{L \times n} \rightarrow \FF_2^*$ (where $\FF_2^*$ denotes the set of finite length binary sequences),  
such that $X(\{W_f : f \in \Fc\},\Fm)$ is the transmitted codeword.  We denote by $J(\{W_f : f \in \Fc\},\Fm)$
the length function (in binary symbols) associated to the encoding function $X$. Each user receives $X(\{W_f : f \in \Fc\},\Fm)$ through the noiseless shared link, and decodes its requested file $W_{f_{u,l}}$, $l=1, \cdots, L$, as 
$(\widehat{W}_{f_{u,1}}, \widehat{W}_{f_{u,2}}, \cdots, \widehat{W}_{f_{u,L}})= \lambda_u(X,Z_u,\Fm)$, 
where  $\lambda_u : \FF_2^* \times \FF_2^{MF} \times \Fc^{L \times n} \rightarrow \FF_2^{LF}$
denotes the decoding function of user $u$. The concatenation of 1) demand vector generation, 2) multicast encoding and transmission over the shared link, and 3) decoding, 
is referred to as the {delivery phase}. 
\hfill $\lozenge$
\end{defn}

We refer to the overall content distribution scheme, formed by both caching and delivery phases, directly as a {\em caching scheme}, and measure the system performance in terms of the rate during the delivery phase. In particular, we define the rate of the scheme as
\begin{equation} \label{average-rate}
R^{(F)} = \sup_{\{W_f : f \in \Fc\}} \; \frac{\EE[ J(\{W_f : f \in \Fc\},\Fsf)]}{F},
\end{equation} 
where the expectation is with respect to the random request vector.\footnote{Throughout this paper, we directly use ``rate" to refer to the average rate defined by \eqref{average-rate} and explicitly use ``average (expected) rate" if needed for clarity.}

Consider a sequence of caching schemes defined by cache encoding functions $\{Z_u\}$,  
multicast coding function $X$, and decoding functions $\{\lambda_u\}$, for increasing file size $F = 1,2,3,\ldots$. 
For each $F$, the worst-case (over the file library) probability of error of the corresponding caching scheme is defined as
\begin{eqnarray} \label{perr}
&& P_e^{(F)} (\{Z_u\},X, \{\lambda_u\})= \notag\\
&& \sup_{\{W_f : f \in \Fc\}} \PP \left( \bigcup_{u \in \Uc} \Big \{ \lambda_u(X,Z_u,\Fsf) \right. \notag\\
&& \quad\quad\quad\quad\quad\quad \left. \neq ({W}_{\fsf_{u,1}}, \cdots, {W}_{\fsf_{u,L}})  \Big \} \right). 
\end{eqnarray}
A sequence of caching schemes is called {\em admissible} if  $\lim_{F \rightarrow \infty} P_e^{(F)} (\{Z_u\},X, \{\lambda_u\}) = 0$. 
{\em Achievability} for our system is defined as follows:

\begin{defn} \label{def:achievable-rate}
A rate $R(n,m,M,L,\qv)$ is {\em achievable} for the shared link caching network with $n$ users, library size $m$, cache capacity $M$, number of requests $L$, 
and demand distribution $\qv$, if there exists a sequence of admissible caching schemes with rate $R^{(F)}$ such that 
\be
\label{eq: Rate}
\limsup_{F \rightarrow \infty} R^{(F)} \leq  R(n,m,M,L,\qv). 
\ee
\hfill $\lozenge$ 
\end{defn}

We let $R^*(n,m,M,\qv)$ denote the infimum (over all caching schemes) of the achievable rates. 
The notion of ``order-optimality'' for our system is defined as follows: 
{ 
\begin{defn}
\label{def: order-optimal}
Let $n,M,L$ be functions of $m$, such that $\lim_{m \rightarrow \infty} n(m) = \infty$. 
A sequence of caching schemes for the shared link caching network with $n$ users, 
library size $m$, cache capacity $M$, number of requests $L$, and demand distribution $\qv$,
is order-optimal if its rate $R(n,m,M,L,\qv)$ satisfies  
\begin{eqnarray}
\limsup_{m \rightarrow \infty} \frac{R(n,m,M,L,\qv)}{R^*(n,m,M,L,\qv)} \leq \nu,
\end{eqnarray}
for some constant $1 \leq \nu < \infty$, independent of $m,n,M$.
\hfill $\lozenge$
\end{defn}
}

\section{Achievability}

In this section, we present an achievable caching scheme based on random popularity-based vector caching and index coding delivery. 

\subsection{RAndom Popularity-based (RAP) Caching}
\label{sec: Caching Placement Scheme}

As in \cite{ji2014average,ji2015random}, we partition each file into $B$ equal-size packets, represented as symbols of $\FF_{2^{F/B}}$, where $F/B$ is sufficiently large (see later).
Let $\Cm$ and $\Wm$ denote the realizations of the {\em packet level}
caching and demand configurations, respectively,
where $\Cm_{u,f}$ denotes the packets of file $f$ cached
at user $u$, and $\Wm_{u,f}$ denotes the packets of file $f$ requested by user $u$. 
We let each user fill its cache independently 
by knowing the caching distribution $\pv=[p_{f}]$, $f\in\{1, \dots, m\}$, with $\sum_{f=1}^m p_{f}=1, \forall u$ and $0 \leq p_{f}\leq1/M, \forall f$. The caching placement is shown in Algorithm \ref{alg1}. 

\begin{algorithm}
\caption{Random Popularity-Based Caching (RAP)}
\label{alg1}
{\small
\begin{algorithmic}[1]
\FORALL{$f \in \mathcal{F}$}
\STATE Each user $u$ caches a subset ($\Cm_{u,f}$) of $p_{f} MB$ distinct packets of file $f$ uniformly at random. 
\ENDFOR
\RETURN $\Cm = [\Cm_{u,f}], u\in\{1, \dots, n\}, f\in\{1, \dots, m\}$.
\end{algorithmic}
}
\end{algorithm}

\subsection{Coded Multicast Delivery}
\label{sec: Coded Transmission}

The coded multicast delivery scheme is based on local chromatic number index coding \cite{shanmugam2013local, ji2014order2}. 
The directed 
conflict graph $\mathcal H_{\Cm, \Wm} = (\Vc, \Ec)$ 
is constructed as follows:
\begin{itemize}
\item Consider each packet requested by each user as a distinct vertex in $\mathcal H_{\Cm, \Wm}=(\Vc, \Ec)$. Hence, each vertex $v\in\Vc$ is uniquely  identified by the pair 
$\{ \rho(v),\mu(v)\}$, where $\rho(v)$ indicates the  \mbox{packet identity} associated to the vertex and 
$\mu(v)$ represents the \mbox{user requesting it}.
\item For any pair of vertices $v_1$, $v_2$, we say that vertex (packet) $v_1$ interferes with vertex $v_2$ 
if the packet associated to the vertex $v_1$, $\rho(v_1)$, is not  in the cache of the user associated to vertex  $v_2$, $\mu(v_2)$, 
and $\rho(v_1)$ and $\rho(v_2)$ do not represent the same packet. Then, 
draw a directed edge from vertex $v_2$ to vertex $v_1$ if $v_1$ interferes with $v_2$.
\end{itemize}

We focus on encoding functions of the following form: for the request vectors $\fv_{u}, u \in \Uc$, the multicast codeword is given by
\be  \label{VV}
X_{\{\fv_{u}, u \in \Uc\}} = \sum_{v \in \mathcal{V}} \omega_v \gv_v = \Gm \omegav,
\ee
where  $\omega_v$ is the binary vector corresponding to packet $v$, represented as a (scalar) symbol of
the extension field $\FF_{2^{F/B}}$, the $\nu$-dimensional vector $\gv_v \in \FF_{2^{F/B}}^\nu$ is the coding vector of packet $\rho(v)$ and where
we let $\Gm = [\gv_1, \dots \gv_{|\mathcal{V}|}]$ and $\omegav= [\omega_1 ,\dots, \omega_{|\mathcal{V}|}]^{\transp}$. 
The number of { columns } $\nu$ of $\Gm$ yields the number of packet transmissions. Hence, the 
transmission rate is given by $\nu/B$ file units.  
To find the desired $\nu$, we introduce the following definition:
\begin{defn} ({\bf Local Chromatic Number})
The directed local chromatic number of a directed graph $\mathcal{H}^d$ is defined as: 
\begin{align}
& \chi_{\rm lc}(\mathcal H^d) = \min_{\cv \in \mathcal{C}} \max_{v\in\mathcal{V}}|\cv(\mathcal{N}^+(v))|
\end{align}
where $\mathcal{C}$ denotes the set of all vertex-colorings of $\mathcal{H}$,  with  $\mathcal{H}$ indicating the  undirected version of $\mathcal H^d$,
$\mathcal{V}$ denotes the vertices of $\mathcal{H}^d$, $\mathcal{N}^+(v)$ is the closed out-neighborhood of 
vertex $v$,\footnote{The closed out-neighborhood of vertex $v$ includes vertex $v$ and all the connected vertices 
via out-going edges of $v$.} 
and $|\cv(\mathcal{N}^+(v))|$ is the total number of colors in $\mathcal{N}^+(v)$ for the given coloring $c$. 
\hfill $\lozenge$
\end{defn}

Given the local chromatic number, for sufficiently large $F/B$, there exists a generator matrix $\Am =  [\av_1, \dots \av_{|\mathcal{C}|}]$ of a $(|\mathcal{C}|, |\chi_{\rm lc}(\mathcal H^d)|)$ MDS code. For all $v \in \Vc$ with the same color, we let $\gv_v = \av_j$ for some $j=1, \cdots, |\mathcal{C}|$. If $v$ and $v'$ are assigned different colors, then $\gv_v \neq \gv_{v'}$. 

To see the decodability of this index code, we introduce the following lemma. 
\begin{lemma}
\label{lemma: full rank}
Let matrix $\Em$ have the following structure,
\be
\Em = \left[\begin{array}{c} \ev_1 \\ \vdots \\ \ev_{|\mathcal{V}|-\chi_{\rm lc}(\mathcal H^d)} \end{array}\right],
\ee
where $\ev_i = [\cdots, 0, 1, 0, \cdots]$, $i = 1, \cdots, |\mathcal{V}|-\chi_{\rm lc}(\mathcal H^d)$, and the position of $1$ can be arbitrary. 
Then the matrix $\left[\begin{array}{c} \Am \\ \Em \end{array} \right]$ is full rank.
\hfill  $\square$
\end{lemma}

From Lemma \ref{lemma: full rank}, letting $\Em$ represent the cached information, it follows that the index code based on the local chromatic number is decodable. 
It can be seen that the index code length is equal to the local chromatic number $|\chi_{\rm lc}(\mathcal H^d)|$ and the index code rate is given by $\chi_{\rm lc}(\mathcal H^d)/B$.   
We refer to this coding scheme as LCIC (local chromatic index coding).\footnote{Instead of using the local chromatic number it is also straightforward to use the fractional local chromatic number to design the coding vector $\Gc$ as illustrated in \cite{shanmugam2013local , ji2014order2}.}
\subsection{Achievable Rate}

It is well known that there is no close form expression of the local chromatic number and that charactering the local chromatic number of a given graph is NP-complete. Hence, in order to analytically approximate its performance, we resort to a greedy local coloring algorithm, named Greedy Constrained Local Coloring (GCLC), introduced in \cite{ji2015Efficient}.\footnote{Due to space limitations, we do not describe GCLC in this paper. The interested reader can refer to \cite{ji2015Efficient} for details.} 
It can be seen that $R^{\rm RAP-LCIC}(n,m,M,L,\qv,\pv) \leq R^{\rm RAP-GCLC}(n,m,M,L,\qv,\pv)$. In the following, our goal is to show that RAP-GCLC is order optimal, which would imply the order-optimality of RAP-GCLC. The achievable rate of RAP-GCLC is given by the following theorem. 
\begin{theorem}
\label{thm: RAP-lCIC}
For the shared link caching network with $n$ users, library size $m$, storage capacity $M$, number of requests $L$ and demand
distribution $\qv$, fix a caching distribution $\pv$. Then, for all $\varepsilon>0$,
\begin{align} 
\label{eq:2}
\lim_{B \rightarrow \infty} &\PP\left(R^{\rm RAP-GCLC}(n,m,M,L,\qv,\pv) \right. \notag\\
&\leq  \left. \min\{\psi(\qv,\pv),\bar m - \bar M\} + \varepsilon\right) = 1
\end{align}

In (\ref{eq:2}), 
\be
\label{eq: m bar simplified}
\bar m = \sum_{f=1}^m \left(1 - \left(1 - q_f\right)^{nL} \right),
\ee
\be 
\label{eq: M bar simplified}
\bar M = \sum_{f=1}^m p_{f} \left(1 - \left(1 - q_{f}\right)^{nL} \right),
\ee
and
\begin{eqnarray}
\label{eq: psi simplified}
&&\psi(\qv,\pv) = \notag\\
&&L\sum_{\ell=1}^n {n \choose \ell}  \sum_{f=1}^m \rho_{f,\ell} (1-p_f M)^{n-\ell+1} (p_f M)^{\ell-1},
\label{eq:3}
\end{eqnarray}
where
$\displaystyle \rho_{f,\ell} \eqdef \mathbb \PP(f = \arg\! \max_{j \in \Dc} \,\,\, (p_jM)^{\ell-1}(1-p_jM)^{n-\ell+1})$
denotes the probability that file $f$ is the file whose $p_f$ maximizes the term $\left((p_jM)^{\ell-1}(1-p_jM)^{n-\ell+1}\right)$ among $\Dc$, which is a set of $\ell$ i.i.d. demands distributed as $\qv$. 
\hfill  $\square$
\end{theorem}

As shown in \cite{ji2015random}, $ \rho_{f,\ell}$ is easy to evaluate. In the following sections, we first evaluate the rate achieved by RAP-GCLC when $\qv$ follows a  
Zipf distribution \cite{breslau1999web} defined as follows: a file $f = 1, \ldots, m$ is requested with probability 
\be
\label{eq: Zipf}
q_f = \frac{f^{-\alpha}}{\sum_{i=1}^m i^{-\alpha}}, \, \forall f = \{1, \cdots, m\},
\ee
where $\alpha \geq 0$ is the Zipf parameter.  
Using the explicit expression for $R^{\rm RAP-GCLC}(n,m,M,L,\qv,\pv)$ in Theorem \ref{thm: RAP-lCIC}, we can optimize the 
caching distribution in order to minimize the number of transmissions. 
We use $\pv^*$ to denote the caching distribution that minimizes $R^{\rm RAP-GCLC}(n,m,M,L,\qv,\pv)$. 
We show that $R^{\rm RAP-GCLC}(n,m,M,L,\qv,\pv^*)$ is order optimal when $\alpha = [0,1)$. 

\section{Uniform Demand Distribution ($\alpha = 0$)}

When $\alpha=0$, $\qv$ follows a uniform distribution. In this case, $p^*_f = \frac{1}{m}, f \in \Fc$ and we refer to this caching distribution as Uniform Placement (UP) and to the corresponding RAP-GCLC  directly as UP-GCLC. 
The achievable rate of UP-GCLC is given by the following theorem. 
\begin{theorem}
\label{theorem: achievable decentralized multiple requests}
For the shared link caching network with $n$ users, library size $m$, storage capacity $M$, number of requests $L$ and random requests following a Zipf distribution $\qv$ with parameter $\alpha=0$, UP-GCLC yields order-optimal rate. 
The corresponding (order-optimal)  rate upper bound is given by
\begin{eqnarray}
\label{eq: upper bound 1}
&& R^{\rm UP-GCLC}(n,m,M,L,\qv,\pv) \notag\\
&&\leq \min\left\{L\left(\frac{m}{M} - 1\right), Ln, m-M\right\}.
\end{eqnarray}
\hfill  $\square$
\end{theorem}

To evaluate the optimality of UP-GCLC, we follow a similar approach as in \cite{ji2014order2, ji2015random} and obtain the following theorem. 
\begin{theorem}
\label{theorem: gap decentralized}
As $n,m \rightarrow \infty$, and $M \leq (1-\varepsilon)m$ for an arbitrarily small positive constant $\varepsilon$, the multiplicative gap between the rate achievable, as  $B \rightarrow \infty$, by UP-GCLC,  $R^{\rm UP-GCLC}(n,m,M,L,\qv,\pv)$, and the lower bound $R^{\rm lb}(n,m,M,L,\qv,\pv)$ is given by Table~\ref{table: 1} and Table \ref{table: 2}, where $c_1 = (1-e^{-1}-\varepsilon')(1-e^{\frac{1}{e}-1}-\varepsilon')$, $c_2 = (1-e^{-1})^2$ and $\varepsilon'>0$ is an arbitrarily small constant.\footnote{Due to the space limitations, we do not present the lower bound $R^{\rm lb}(n,m,M,L,\qv,\pv)$ in this paper. This converse can be shown by combing the ideas in \cite{ji2014order2, ji2015random}.}
\begin{table}
\centerline{\includegraphics[width=7cm]{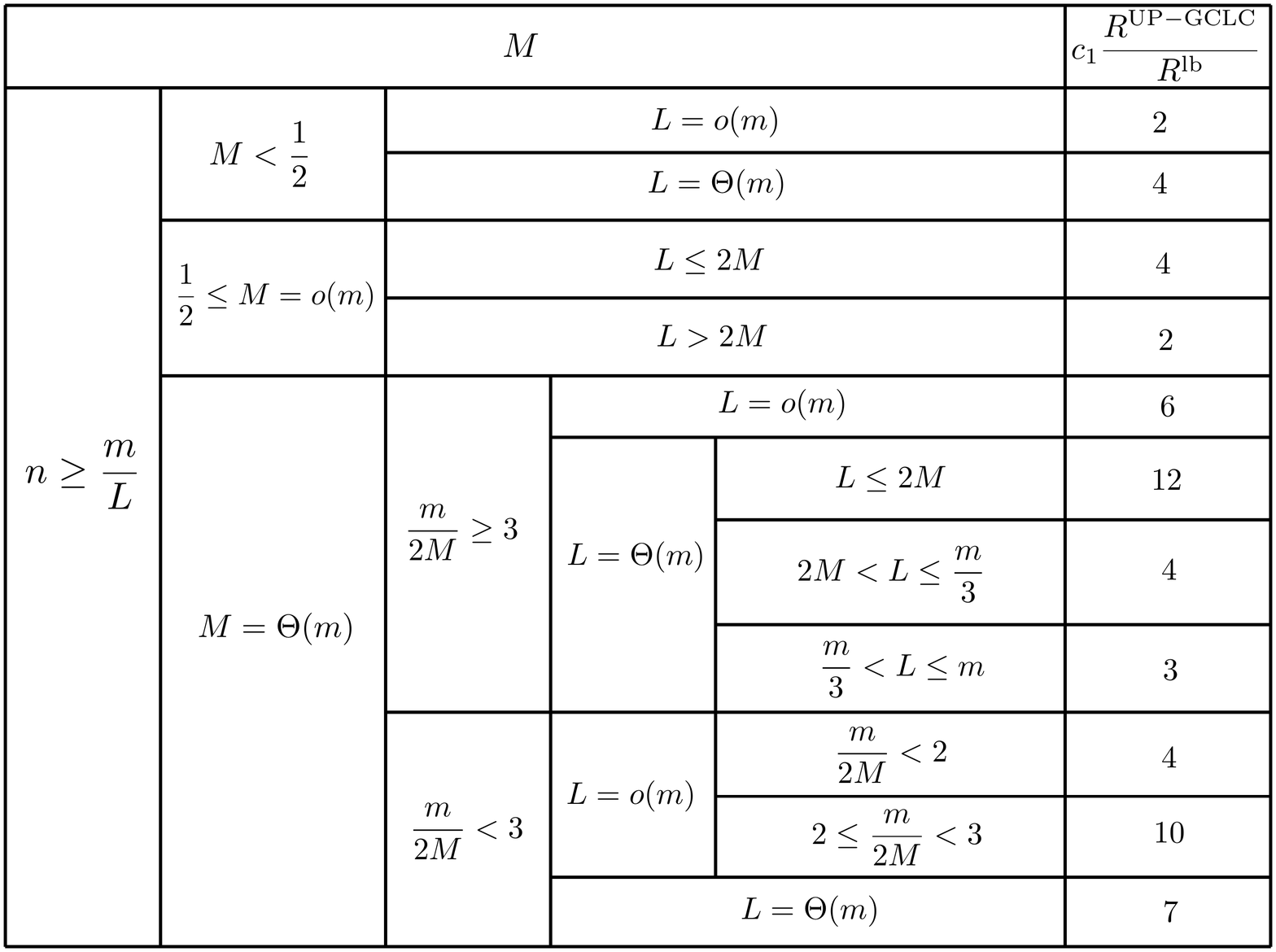}}
\vspace{-0.2cm}
\caption{When $n \geq \frac{m}{L}$, this table this table shows the upper bound of $\frac{R^{\rm UP-GCLC}(n,m,M,L,\qv,\pv)}{R^{\rm lb}(n,m,M,L,\qv,\pv)}$, which is denoted as $\frac{R^{\rm UP-GCLC}}{R^{\rm lb}}$.}
\vspace{-0.3cm}
\label{table: 1}
\end{table}
\begin{table}
\centerline{\includegraphics[width=7cm]{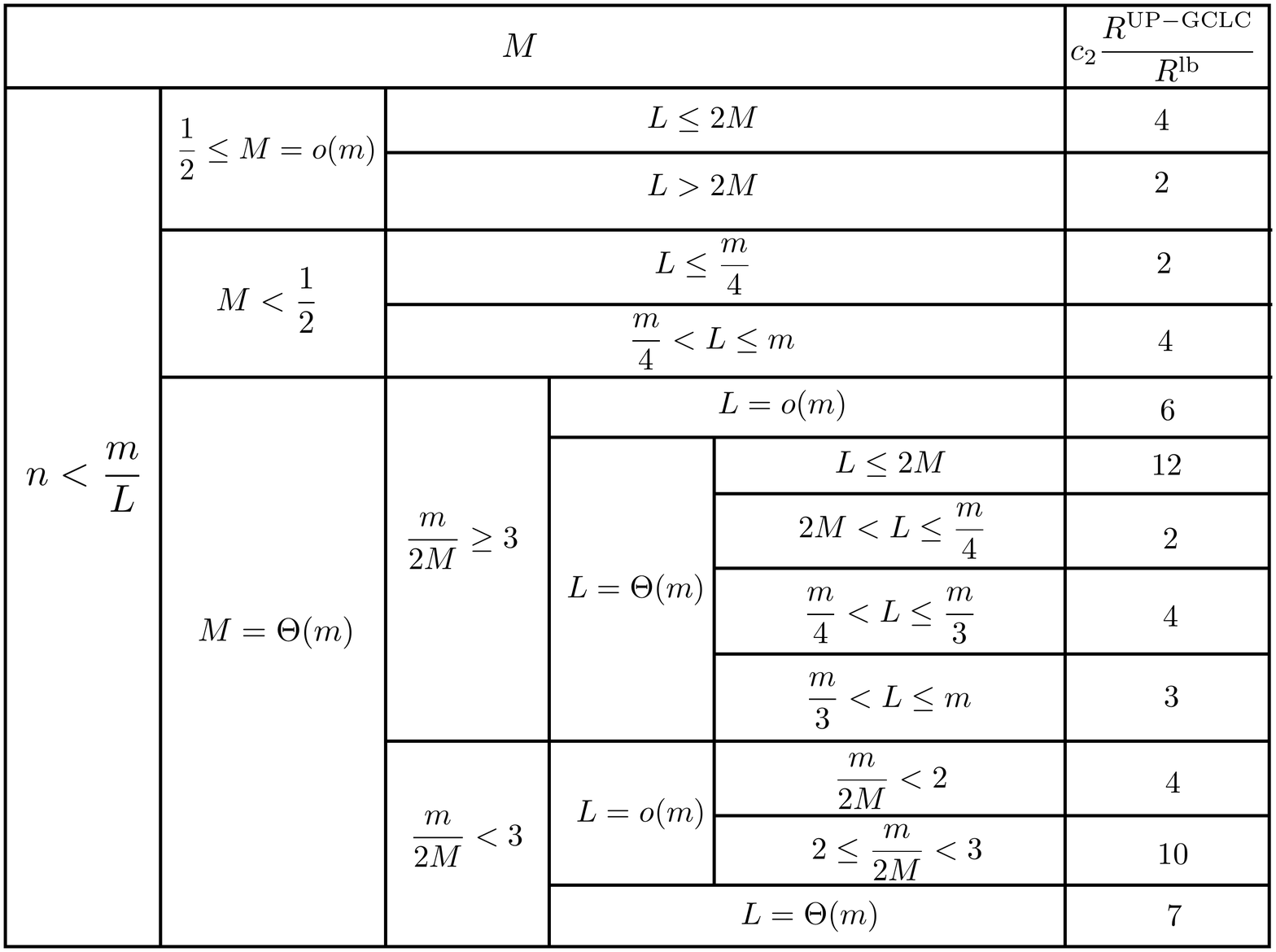}}
\vspace{-0.2cm}
\caption{When $n < \frac{m}{L}$, this table this table shows the upper bound of $\frac{R^{\rm UP-GCLC}(n,m,M,L,\qv,\pv)}{R^{\rm lb}(n,m,M,L,\qv,\pv)}$, which is denoted as $\frac{R^{\rm UP-GCLC}}{R^{\rm lb}}$.}
\vspace{-0.7cm}
\label{table: 2}
\end{table}
\hfill  $\square$
\end{theorem}


\section{Nonuniform Demand Distribution  ($\alpha \in (0,1)$)}

When $\alpha \in (0,1)$, evaluating  $\pv^*$  becomes quite difficult.
To this end,  as in \cite{ji2015random},  we approximate $\pv^*$ via a {\em truncated uniform distribution} $\widetilde \pv$, 
referred to as the Random Least Frequently Used (RFLU) caching distribution, 
where $p_f = \frac{1}{\widetilde m}$ for $f=1,\cdots,\widetilde m$; $p_f = 0$ for $f=\widetilde m+1, \cdots, m$,
and where the cut-off index $\widetilde m \geq M$ is a function of the system parameters.  
Note that when $\widetilde m = m$,  RLFU becomes UP. 
By using Theorem \ref{thm: RAP-lCIC}, the achievable rate of RLFU-GCLC is given by: 
\begin{corollary}
For the shared link caching network with $n$ users, library size $m$, storage capacity $M$, number of requests $L$, and Zipf demand distribution $\qv$
with parameter $\alpha>0$, the rate achieved by RLFU-GCLC is upper bounded by:
\begin{eqnarray}
\label{eq: upper bound 2}
&& R^{\rm RLFU-GCLC}(n,m,M,L,\qv, \tilde \pv)  \notag\\
&&\leq \min\left\{\min\left\{L\left(\frac{\widetilde m}{M} - 1\right), \widetilde m - M\right\} \right. \notag\\
&&\quad \left. +\min\{Ln(1-G_{\widetilde m}), m-\widetilde m\}, Ln, m-M\right\}, \notag 
\end{eqnarray}
where $G_{\widetilde m} = \sum_{f=1}^{\widetilde m}q_f$. 
\hfill  $\square$
\end{corollary}

\begin{figure}[ht]
\centerline{\includegraphics[width=6cm]{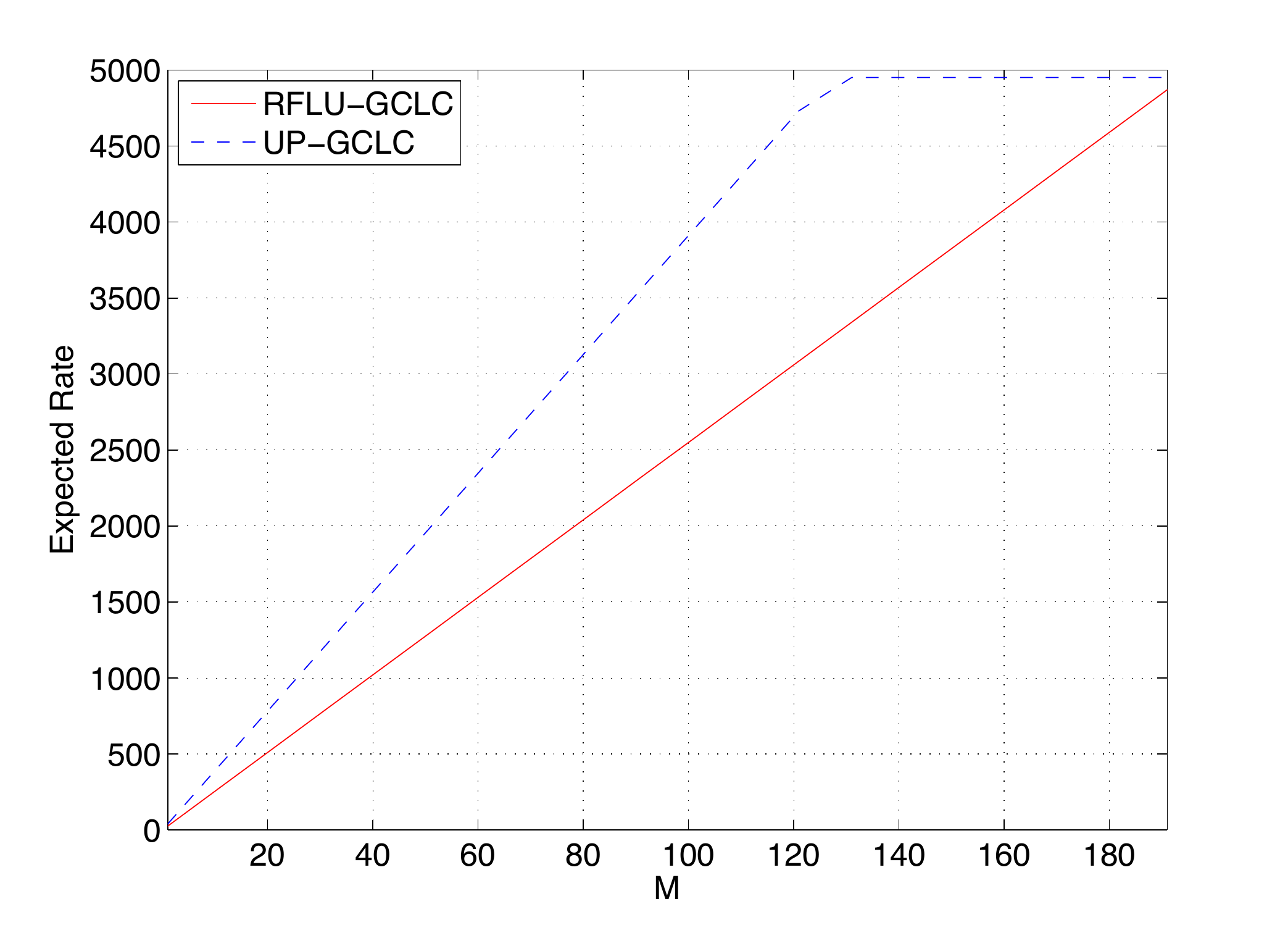}}
\vspace{-0.4cm}
\caption{The upper bound of the achievable rate by RFLU-GCLC and UP-GCLC, where $n=50,m=5000,M=50$ and $\alpha=0.9$.}
\vspace{-0.2cm}
\label{fig: RFLU-GCLC}
\end{figure}

Note that for $\alpha\neq 0$, choosing the $\widetilde{m}$ that  minimizes $R^{\rm RLFU-GCLC}(n,m,M,L,\qv, \tilde \pv) $ 
can provide a significant gain compared to UP ($\widetilde{m}=m$), as shown in Fig. \ref{fig: RFLU-GCLC}.  
From Fig. \ref{fig: RFLU-GCLC}, we can see that when $n=50,m=5000,M=50$ and $\alpha=0.9$, RLFU-GCLC can reduce by approximately $30\%$ the rate of UP-GCLC.
However, in terms of order-optimality, interestingly, even when $\alpha \in (0,1)$, uniform caching placement is sufficient to achieve the order-optimal rate,\footnote{This is due to the heavy tail property of the Zipf distribution when $\alpha \in (0,1)$. In fact, for $\alpha \in (0,1)$, as $m \rightarrow \infty$, the probability mass is ``all in the tail'', i.e., the probability $\sum_{f=1}^{\widetilde{m}} q_f$ of the most probable $\widetilde{m}$ files vanishes, for any finite $\widetilde{m}$.} 
as stated in the following theorem.

\begin{theorem}
\label{theorem: nonuniform multiple requests}
For the shared link caching network with $n$ users, library size $m$, storage capacity $M$, number of requests $L$ and random requests following a Zipf distribution $\qv$ with parameter $0 < \alpha<1$, let $M \leq (1-\varepsilon)m$ for an arbitrarily small positive constant $\varepsilon$, as  $B \rightarrow \infty$, UP-GCLC yields order-optimal rate. 
The corresponding (order-optimal) rate upper bound is given by (\ref{eq: upper bound 1}).
\hfill  $\square$
\end{theorem}



\section{Scalar Random Caching Placement}

Recall that to achieve the order optimal rate of RAP-GCLC, we need to let $B$ be sufficiently large, which may not be feasible in practice. In fact, \cite{shanmugam2014} showed that when $L=1$ and users make distinct requests (worst-case scenario), it is necessary that $B$ grows exponentially with $n$ under UP. Intuitively, this is because a sufficiently large $B$ can help to create coding opportunities such that one transmission can benefit multiple users. However, unlike the single request case, when each user makes multiple requests, the number of requests $L$ can play a similar role as $B$ for the single request case when $M$ is large enough. In other words, if each user caches entire files, a sufficiently large $L$ can also help create coding opportunities.  

Let the caching phase be given by a Scalar Uniform Placement (SUP) scheme, 
in which each user caches $M$ entire files chosen uniformly at random.
\footnote{For simplicity, we let $M$ be an positive integer.} 
For the delivery phase, GCLC is applied. Then, 
letting $L \rightarrow \infty$ as a function of $n,m,M$, we obtain the following theorem.  
\begin{theorem}
\label{theorem: B equal 1}
For the shared link caching network with $n$ users, library size $m$, storage capacity $M = \omega\left(L\right)$, and $L$ distinct per-user requests, 
when $n,m,M,L \rightarrow \infty$ as:  $L = \omega\left(\frac{nM}{m}\left(\frac{m}{M}\right)^{n} \frac{1}{\left(1-\frac{M}{m}\right)}\right)$ when $\frac{M}{m} < \frac{1}{2}$;  $L = \omega\left(n \left(\frac{m}{m-M}\right)^n\right)$ when $\frac{M}{m} \geq \frac{1}{2}$; 
then the achievable rate of SUP-GCLC ($B=1$) is upper bound by
\begin{eqnarray}
\label{eq: upper bound 3}
&& R^{\rm SUP-GCLC}(n,m,M,L,\qv,\pv) \notag\\
&&\leq (1+o(1))\min\left\{L\left(\frac{m}{M} - 1\right), Ln, m-M\right\}. \notag
\end{eqnarray}
\hfill  $\square$
\end{theorem}

From Theorems \ref{theorem: achievable decentralized multiple requests} and \ref{theorem: B equal 1}, we can see that when $L$ and $M$ are large enough,  instead of requiring a large $B$  and vector (packet-level) coding, a simpler scalar (file-level) coding scheme is sufficient to achieve the same order-optimal rate. We remark, however, that the range of the parameter regimes in which this result holds is limited due to the requirement of a large $M$ and $L$, and hence an exact equivalence between the effect of $B$ and $L$ on the order-optimal rate does not hold in general. 
In practice, it is important to find 
the right balance between $B$ and $L$ given the remaining system parameters. 



\bibliographystyle{IEEEtran}
\bibliography{references_d2d_2,references_2} 

\end{document}